\documentclass[10pt]{article}
\usepackage{amsmath}
\usepackage{amsmath, dsfont}
\usepackage{amssymb}
\usepackage[dvips]{graphics}
\usepackage[dvipsnames]{xcolor}
\usepackage{epsfig}
\usepackage{calc}
\usepackage{amsfonts}
\usepackage{graphicx}
\usepackage[ansinew]{inputenc}
\usepackage{color}

\newcommand{\be}{\begin{equation}}
\newcommand{\ee}{\end{equation}}
\newcommand{\bea}{\begin{eqnarray}}
\newcommand{\eea}{\end{eqnarray}}
\newcommand{\beal}{\begin{aligned}}
\newcommand{\eeal}{\end{aligned}}

\newcommand\B{{\mathcal{B}}}
\newcommand{\HH}{{\mathcal{H}}}

\newcommand{\ZZ}{{\mathcal{Z}}}

\title{Regular black holes and gravitational particle-like solutions in generic DHOST theories}

\author{Olaf Baake$^{\dag\ddag}$, Christos Charmousis$^{\sharp}$, Mokhtar Hassaine${^\dag}$ and  Miguel {San Juan}${^\dag}$\\
\normalsize{$^{\dag}$Instituto de
Matem\'aticas, Universidad de Talca, Casilla 747, Talca 3460000,
Chile,} \\ \normalsize{$^{\ddag}$Centro de Estudios Cientificos (CECs), Av.
Arturo Part 514, Valdivia 5090000, Chile},\\
\normalsize{$^{\sharp}$Université Paris-Saclay, CNRS/IN2P3, IJCLab, 91405 Orsay, France.}}
\let\ssection=\section
\renewcommand{\section}{\setcounter{equation}{0}\ssection}
\date{}

\begin{document}
\maketitle
\begin{abstract}
We construct regular, asymptotically flat black holes of higher order scalar tensor (DHOST) theories, which are obtained by making use of a generalized Kerr-Schild solution generating method. The solutions depend on a mass integration constant, admit a smooth core of chosen regularity, and generically have an inner and outer event horizon. In particular, below a certain mass threshold, we find  massive, horizonless, particle-like solutions. We scan through possible observational signatures ranging from weak to strong gravity and study the thermodynamics of our regular solutions, comparing them, when possible, to General Relativity black holes and their thermodynamic laws.

\end{abstract}

\section{Introduction}

It is an undeniable fact, whose origin goes back to the Schwarzschild
solution, that the notion of a black hole is intimately linked to the
concept of spacetime singularities. In fact, it is well-known that,
under certain energy conditions, classical solutions of general
relativity exhibit singularities as a direct
consequence of the so-called singularity theorems
\cite{Penrose:1964wq, Hawking:1969sw}. The appearance of
singularities is essentially due to the classical character of the
theory of general relativity, and a quantum theory of
gravity may be expected to cure such pathologies.

However, in the absence of a complete theory of quantum gravity one can search for black hole spacetimes with a global structure similar to the well-known solutions (like the Schwarzschild or Reissner-Nordström solutions), but
in which the central singularity is absent. Such solutions are commonly known as regular black holes. These ideas originate from the pioneering works of Sakharov \cite{Sakharov:1966aja}, Gilner \cite{1966JETP...22..378G} and also Bardeen \cite{bardeen1968non} who presented the first
example of a regular black hole as an ad-hoc metric (not originating from an action). A physical construction of the Bardeen metric as a solution of a given action was finally proposed much later in \cite{AyonBeato:2000zs}. There the authors showed
that the Bardeen metric can be obtained from the Einstein
equations with a non-linear magnetic source. Although the Bardeen
metric was the first example of a regular spacetime, the first exact
regular black hole solution of a given theory was found by Ay\'on-Beato and Garcia
\cite{AyonBeato:1998ub} for the Einstein equations coupled to a specific non-linear electrodynamic source.

Models involving non-linear electrodynamics have been fruitful in the construction of regular solutions, see e.g. Refs. \cite{AyonBeato:1999rg}, and also Ref. \cite{Ansoldi:2008jw} for a review. Many of these regular black holes present a de Sitter core at the origin, and their regularity is quantified by a regularizing parameter identified with a non-linear electrodynamic charge. It is also important to stress that the parameter of regularization is not a constant of integration but is rather
an input of the matter action. This observation has important consequences, for example
on the thermodynamic properties of these regular solutions. Indeed, depending on whether the regularizing parameter is considered as a varying parameter or not, the thermodynamic properties may be different. In order to illustrate this fact, one can note that for the Bardeen regular black hole, the one-quarter area law of the entropy is usually violated \cite{Sharif:2010pj} when considering a non varying  magnetic charge, while this "universal" law can be restored by promoting the magnetic charge as a variable parameter \cite{Ma:2014qma}. Note that in certain non-minimally coupled Lagrangians analytic regular solutions were found where the mass and the charge truly are integration constants \cite{Cano:2020ezi}.

In this work we will construct regular black hole solutions, which are asymptotically very similar to Schwarzschild, without the need of introducing an additional regularization parameter inherent to the action. For these black holes, regularity will not be enforced by the fine tuning of some action parameter, it will rather be achieved due to the functional form of the regularizing function appearing in the solutions. In other words,
the fall-off of the mass term of our solutions turns out to be an analytic function with a de Sitter core at the origin as a consequence of the field equations.
The degree of regularity and its strength are monitored by two parameters, one fixing the core to be de Sitter or higher and one fixing the strength of the higher order term against the mass of the black hole. The regular black holes found here are exact solutions of scalar tensor theories beyond those initially proposed by Horndeski \cite{Horndeski:1974wa}. The regularizing function sets, as one would expect, the scalar degree of freedom without any fine tuning of the theory.

The scalar tensor theories have higher than second derivative equations of motion and are (still) free of Ostrogradski type pathologies \cite{Langlois:2015cwa,Crisostomi:2016tcp}. These general Lagrangians have been dubbed {\it Degenerate Higher Order Scalar Tensor} (DHOST) or {\it Extended Scalar Tensor} (EST) theories \cite{Langlois:2015cwa,Crisostomi:2016tcp,Motohashi:2016ftl}, and are widely studied in the literature (see for example \cite{Babichev:2017guv, Babichev:2017lmw, Charmousis:2019vnf, Babichev:2016kdt} for their compact objects and \cite{Kobayashi:2019hrl} for a review). More precisely, we will consider the following class of shift symmetric and parity preserving DHOST theories that contain up to second order covariant derivatives of the scalar field (in the action),
\begin{eqnarray}
S[g,\phi]=\int
d^4x\sqrt{-g}&\Big[K(X)+G(X)R+A_1(X)\left[\phi_{\mu\nu}\phi^{\mu\nu}-(\Box\phi)^2\right]+A_3(X)\Box\phi\,\phi^{\mu}\phi_{\mu\nu}\phi^{\nu}
\nonumber\\
&+A_4(X)\phi^{\mu}\phi_{\mu\nu}\phi^{\nu\rho}\phi_{\rho}+A_5(X)\left(\phi^{\mu}\phi_{\mu\nu}\phi^{\nu}\right)^2\Big],
\label{action}
\end{eqnarray}
where the coupling functions $K, G, A_1, A_3, A_4$ and $A_5$ depend
only on the kinetic term of the scalar field
$X=g^{\mu\nu}\phi_{\mu}\phi_{\nu}$, and where
$\phi_{\mu}=\partial_{\mu}\phi$ and
$\phi_{\mu\nu}=\nabla_{\mu}\nabla_{\nu}\phi$. The coupling functions
$A_4$ and $A_5$ are chosen to satisfy
\begin{eqnarray}
\label{a4a5} A_4&=&\frac{1}{8(G-XA_1)^2} \left\{ 4G \left[
3(-A_1+2G_{X})^2-2A_3G\right] -A_3X^2(16A_1G_{X}+A_3 G) \right.
\nonumber\\
&& \left. + 4X \left[
-3A_2A_3G+16A_1^2G_{X}-16A_1G_{X}^2-4A_1^3+2A_3G G_{X} \right]
\right\},
\nonumber\\
A_5&=&\frac{1}{8(G-X A_1)^2} (2A_1-XA_3-4G_{X})
\left(A_1(2A_1+3XA_3-4G_{X})-4A_3G\right)
\end{eqnarray}
in order to ensure  the absence of Ostrogradski ghosts \cite{Langlois:2015cwa, Crisostomi:2016tcp}. Recently, it has been shown that regular black hole solutions for this class of theories can be constructed (including the well known cases of Bardeen \cite{bardeen1968non} or Hayward metrics \cite{Hayward:2005gi}), see Ref. \cite{Babichev:2020qpr}. The algorithm of construction is a byproduct of extending the Kerr-Schild solution generating method to scalar tensor theories. The key point in extending this well known method from GR is the assumption that the kinetic term of the scalar field remains unchanged under the static (usual) Kerr-Schild transformation. Another crucial observation that we make here is that regular black holes cannot belong to Horndeski theory. We will see that the theories involving regular black holes  correspond to a conformal and disformal map originating from Horndeski theory and ultimately belong to a pure DHOST theory. We will trace the reason for this to the recent interesting work discussing singularities in scalar tensor theories \cite{Domenech:2019syf}.

We would like to note that although the kinetic term of the
scalar field will be assumed to be only depending on the radial
coordinate, this does not exclude the fact that the scalar field can
depend linearly, for example, on the time coordinate, i.e. $\phi(t,r)=\alpha
t+\psi(r)$ where $\alpha$ is a constant. This possibility is attributed to the higher order nature of DHOST theory, and the shift invariance symmetry of the scalar field. The scalar time dependence was first used in \cite{Babichev:2013cya} and has been found recently to be related to the geodesics of spacetime \cite{Charmousis:2019vnf} whenever the kinetic term $X$ is constant. In fact, in the case of higher order scalar tensor theories, examples of compact objects with a linear time dependent scalar field have been found, see e.g. \cite{Babichev:2013cya,Kobayashi:2014eva,Bravo-Gaete:2013dca,
Lehebel:2018zga}. In particular, stationary solutions, which are distinctively different from the Kerr spacetime \cite{Anson:2020trg}, have been recently constructed.

In our search for regular black holes we will focus on a static scalar field where $X$ will not be a constant function. This is a crucial requirement as $X$ will also play the role of the regularizing function smoothing out the geometry near the origin. Once we obtain our regular solution we will discuss its most important properties. We will then proceed to study its possible observational characteristics scanning from weaker to stronger gravity effects.

The plan of the paper is organized as follows. In the next section,
we will explicitly write the field equations associated to the
variation of the DHOST action (\ref{action}-\ref{a4a5}). The key steps of the Kerr-Schild solution generating method \cite{Babichev:2020qpr} will also be outlined, in order to explicitly construct a family of regular asymptotically flat black holes, that are solutions of some specific DHOST action (\ref{action}-\ref{a4a5}) with coupling functions
specified in the Appendix. We will analyze the solutions and discuss the leading Post-Newtonian parameters, precession effects and 
null geodesics, scanning through observable signatures. In Sec. $3$, the thermodynamic analysis of these regular solutions will be carried out through the Euclidean method, and we will show that the regularity condition of the solutions is incompatible with the area law of the entropy. In spite
of this, the first law of thermodynamics is shown to hold for the regular solutions. Our conclusions will be presented in Sec. $4$.

\section{Field equations and construction of regular black holes}

We will be dealing with a four-dimensional scalar tensor theory described by the metric $g$ and a single scalar field $\phi$ whose dynamics is governed by the action (\ref{action})
and whose coupling functions $A_4$ and $A_5$ are given by (\ref{a4a5}).
We will focus on static metrics with a scalar field such that its standard kinetic term $X=g^{\mu\nu}\phi_{\mu}\phi_{\nu}$ only
depends on the radial coordinate $r$, i.e.
\begin{eqnarray}
ds^2=-h(r)\,dt^2+\frac{dr^2}{f(r)}+r^2\left(d\theta^2+\sin(\theta)^2
d\varphi^2\right),\qquad X=g^{\mu\nu}\phi_{\mu}\phi_{\nu}:=X(r).
\label{ansatz}
\end{eqnarray}
For this ansatz, the field equations associated with the DHOST action
(\ref{action}-\ref{a4a5}) are conveniently written as
\begin{subequations}
\label{seed3}
\begin{eqnarray}
&&  X[2(A_1 G)_X + G A_3]+r^2\left[(K\HH)_X+\frac{3}{4} K \B\right]=0,\label{seed1}\\
&&-3(\B r X')^2 +8 (\B r X') \HH \left(\frac{rh'}{h} +4\right) - 32 \HH \left[\frac{K r^2 +2  G}{f}+2\HH \left(\frac{rh'}{h} +1\right)\right]=0,\label{seed2}\\
&& r^2(16 \B_X \HH + 3 \B^2) X'^2+8\HH X' r\left(\B r
\frac{f'}{f}-16 \HH_X\right)+16 r^2 \HH \B X'' \nonumber\\ && -64\HH^2\left[\left(\frac{rf'}{f} +1 \right) +\frac{2  G +
r^2 K}{2f\HH}\right]=0,
\label{seed4}
\end{eqnarray}
\end{subequations}
where $(\;)'$ denotes the derivative with respect to the radial
coordinate, $r$, while subscript $X$ denotes the derivation with respect
to the kinetic term $X$. To simplify the notation, we have defined the auxiliary functions of the action,
\begin{eqnarray}
&&\HH(X)=A_1(X)\,X-G(X),\qquad \B(X)=A_3(X)\,X+4G_X(X)-2A_1(X),\nonumber\\
&&\ZZ(X)=A_3(X)+A_4(X)+X\,A_5(X). \label{defHBZ}
\end{eqnarray}

Another interesting note is the Horndeski limit \cite{Horndeski:1974wa} and the beyond Horndeski limit \cite{Gleyzes:2014qga, Gleyzes:2014dya} of our general DHOST theory equations. Indeed, (quartic) Horndeski theory, parameterized by $G_4=G$ is attained with $2 G_X=A_1=-A_2$ and $A_3=0$, while quartic beyond Horndeski is given by $2 G_X-X F=A_1=-A_2$ and $A_3=-2 F$. The function $F$ is the quartic beyond Horndeski term which is in a one to one correspondence with the disformal transformation, mapping Horndeski to beyond Horndeski theory (see for example the nice analysis in \cite{Crisostomi:2016tcp}). In particular, we note that in both cases of quadratic Horndeski and beyond Horndeski we have $\B=0$, which means that $\B$ in our field equations represents the conformal transformation mapping beyond Horndeski to pure DHOST theory. We will come back to this observation in a moment.

In order to be self-contained, we will briefly recall the procedure
described in \cite{Babichev:2020qpr} which allows the construction
of regular black hole solutions from simple seed configurations. The
first step is to look for a simple seed solution of the field
equations (which does not describe a black hole) and schematically
represent it by
\begin{eqnarray}
ds_0^2=-h_0(r)dt^2+\frac{dr^2}{f_0(r)}+r^2\left(d\theta^2+\sin(\theta)^2
d\varphi^2\right),\qquad
X_0=g_{(0)}^{\mu\nu}\phi^{(0)}_{\mu}\phi^{(0)}_{\nu}:=X_0(r).
\label{seedconfi}
\end{eqnarray}
Now, as shown in Ref. \cite{Babichev:2020qpr}, the equations of
motion (\ref{seed3}) are invariant under a Kerr-Schild
transformation of the metric, provided that the kinetic term of
the scalar field is left invariant. More precisely, it is straightforward to see that
the equations (\ref{seed3}) are invariant under the following
simultaneous transformations
\begin{eqnarray}
h_0(r)\to h_0(r)-2\mu \frac{m(r)}{r},\qquad  f_0(r)\to
\frac{f_0(r)}{h_0(r)}\left(h_0(r)-2\mu \frac{m(r)}{r}\right),\quad
\mbox{with}\,\,\,\,\, m(r)=e^{\frac{3}{8}\int
dX\frac{\B(X)}{\HH(X)}}, \label{ks}
\end{eqnarray}
and $X$ remains unchanged, i. e. $X_0(r)=X(r)$. Here $\mu$ is a
constant that will be shown to be proportional to the mass of the resulting solution. Our
second step is to use this Kerr-Schild symmetry (\ref{ks}) to deduce that the configuration given by,
\begin{eqnarray}
&&ds^2=-\left(h_0(r)-2\mu\,\frac{m(r)}{r}\right)dt^2+\frac{h_0(r)\,dr^2}{f_0(r)\left(h_0(r)-2\mu\,\frac{
m(r)}{r}\right)}+r^2\left(d\theta^2+\sin(\theta)^2
d\varphi^2\right),\nonumber\\
&& X(r)=g^{\mu\nu}\phi_{\mu}\phi_{\nu}=X_0(r), \label{ansatzz}
\end{eqnarray}
will satisfy the same equations as those satisfied by the simple
seed solution (\ref{seedconfi}), provided that the mass function $m(r)$ is
given by
\begin{eqnarray}
m(r)=e^{\frac{3}{8}\int dX\frac{\B(X)}{\HH(X)}}.
\label{massfunction}
\end{eqnarray}
Note that in order for the mass term to be non trivial (i.e. with a non-Newtonian fall-off) we need to venture outside of beyond Horndeski theory, where $\B\neq 0$. According to the observation made in the previous paragraph, $\B$ is related to the conformal degree of freedom for pure DHOST theory. This leads us to the conclusion that we must have a combined disformal and conformal transformation of Horndeski theory to have any hope of constructing a regular solution. The regular solutions are crucially situated in higher order DHOST theory-not in Horndeski or beyond Horndeski theory.

To keep things simple we make the following working hypothesis \cite{Babichev:2020qpr}
\begin{equation}
\frac{3\B}{8\HH}=\frac{1}{X} \Longrightarrow
m(r)=X(r), \label{hyp1}
\end{equation}
Hence, starting from a seed metric, the ``choice'' of the
mass function $m(r)$, or equivalently of the seed kinetic term (\ref{hyp1}) will be key in order to ensure the regularity of the final (massive) configuration (\ref{ansatzz}) at the origin and at infinity. Moreover, once we fix the expression of $X_0(r)$ as an invertible function, we will be able to specify the corresponding DHOST theory (\ref{action}-\ref{a4a5}), that is to determine the functions $K, G, A_1$ and $A_3$ (as functions of $X$
only) \cite{Babichev:2020qpr}. For example, in the asymptotically flat case with a seed
metric $f_0=h_0=1$, the regularity at the origin will be ensured if $m(r)=\mathcal{O}(r^3)$.
Indeed, in this case the solution is shown to exhibit a de Sitter core at the origin, ensuring that any invariant constructed out of the Riemann tensor will be regular at the origin.
Given these preliminary requirements we see that it is essential to be in the context of DHOST theory, in order to find regular black holes in accordance with the discussion and findings in \cite{Domenech:2019syf}. Hence, regular black holes are necessarily solutions of a pure DHOST theory. In other words, such regular solutions would be images of the mapping of a combined conformal and disformal transformation of a Horndeski solution.

\subsection{Asymptotically flat regular black holes}

We will first focus on the construction of asymptotically regular
black holes with a flat seed metric given by $h_0=f_0=1$. In this
case, following the results obtained in Ref. \cite{Babichev:2020qpr}, one can easily express $\HH$ and $G$ as
\begin{eqnarray*}
\HH=\frac{1}{X \left( \frac{ r X'}{3X}-1\right)},\qquad G =\frac{1}{ X}\left(1-\frac{
rX'}{X} \right)-\frac{K r^{2}}{2}.
\end{eqnarray*}
Now, in order to get the coupling
function $K$, we first write
\begin{eqnarray}
A_3=-\frac{4G_{X}}{X}+\frac{2A_{1}}{X}+\frac{8\HH}{3X^2},\qquad
A_1=\frac{\HH+G}{X} \label{id1}
\end{eqnarray}
and then inserting the expressions (\ref{id1}) into
Eq.(\ref{seed1}), we obtain, after some algebraic manipulations,
\begin{equation}
2(\HH
G)_X+r^2(K\HH)_X+\frac{2\HH}{X}\left(\frac{4}{3}G+Kr^2\right)=0.
\label{edo}
\end{equation}
Finally, the coupling function  $K$ is shown to be given by {\small
\begin{eqnarray*}
K  =-{\frac {2 \left[3 X \left(r X''+2X'\right)+
r^{2}X^{-1} X'^{3} - 7rX'^{2}
\right]}{r  X \left( r X'-3 X\right)^{2}} }.
\end{eqnarray*}}
We are now ready to construct an explicit family of regular
black hole solutions. We will opt for
a (seed) kinetic term,
\begin{eqnarray}
X(r)=X_0(r)=\frac{2}{\pi}\arctan(\frac{\pi r^p}{2 \sigma^{p-1}}). \label{kinchoice}
\end{eqnarray}
The function $X$ depends on the integer $p$ and the bookkeeping parameter $\sigma$. In particular, the limiting case $\sigma\to 0$ gives us the usual Schwarzschild case.
Our choice is motivated from three essential requirements emanating from the resulting metric function,
$h(r)= 1-\frac{2\mu X(r)}{r}$ :
\begin{itemize}
\item First of all, for $r$ close to the origin we have,
\begin{equation}
h(r)=1-2\mu \left(\frac{r}{\sigma}\right)^{p-1}+O(r^{3p-1}),
\end{equation}
and hence, as shown below for $p\geq 3$, $\sigma\neq 0$
, the final metric will be regular at the origin. The de Sitter core is attained for $p=3$, and increasing regularity from there on for $p>3$. \\
\item Secondly, $X$ asymptotes unity for large $r$, and as such gives for $h$ a similar behavior at asymptotic infinity to the Schwarzschild solution. We have,
\begin{equation}
h(r)=1-\frac{2\mu}{r}+\frac{8\mu\sigma^{p-1}}{\pi^2 r^{p+1}}+O(r^{3p+1}),
\end{equation}
 \item Last but not least, the function $X(r)$ is bijective for our coordinate range $r\in [0,\infty[$.
\end{itemize}
Using the latter property one can see that the seed configuration, $h_0=f_0=1$, with a kinetic term given by (\ref{kinchoice}), is a solution of the DHOST action (\ref{action}-\ref{a4a5}) with coupling functions reported in the Appendix. Crucially, the action functionals are only functions of $X$, and the theory parameters, $\sigma$ and $p$. The power, $p$, fixes the solution's core regularity at the origin. Once $p$ is fixed, the solution is regular without any fine-tuning of the parameter $\sigma$, which has been inserted so as to track down differences from GR at $\sigma\to 0$.
Using therefore the generalized Kerr-Schild transformation, one determines that the solution given by
\begin{eqnarray}
&&ds^2=-\left(1-\frac{4\mu\arctan(\frac{\pi r^p}{2 \sigma^{p-1}})}{r\pi}\right)dt^2+\frac{dr^2}{ \left(1-\frac{4\mu\arctan(\frac{\pi r^p}{2 \sigma^{p-1}})}{r\pi}\right)}+r^2\left(d\theta^2+\sin(\theta)^2
d\varphi^2\right),\nonumber\\
&&X(r)=\frac{2}{\pi}\arctan(\frac{\pi r^p}{2 \sigma^{p-1}}),\label{solreflat}
\end{eqnarray}
satisfies the field equations of the DHOST action (\ref{action}-\ref{a4a5}) with coupling functions given in the Appendix, which has been additionally verified by inserting this solution directly into the equations of motion.

\begin{figure}[h]
\centering
\includegraphics[scale=0.4]{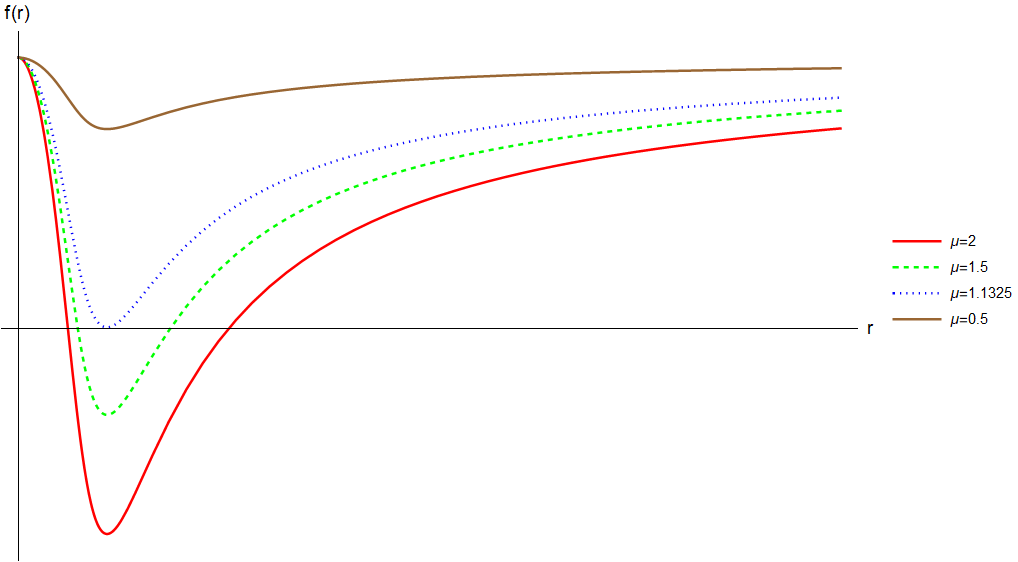}
\caption{Metric function $g_{00}$ for $p=3$ and $2\sigma^2=\pi$. The inner and outer horizons correspond to the roots of the function, while for smaller masses than $\mu_{ext}$ (blue dotted curve) the solution has no horizon.}
\label{metric_function_p3}
\end{figure}

Let us now make some comments on the properties of
(\ref{solreflat}). First of all, for $p>0$, the metric solution will
behave asymptotically $(r\to\infty)$ as the Schwarzschild spacetime.
For $\mu>0$ and $p>0$, the metric solution has an
inner and an outer event horizon as we see from the plot in Fig. \ref{metric_function_p3}. The outer horizon is an event and Killing horizon (for the Killing vector $\partial_t$), which is manifest by preforming the usual Eddington-Finkestein coordinate
transformation. The inner horizon is a Cauchy horizon for any timelike hypersurface situated in the exterior spacetime where $\partial_t$ is timelike. The solution has a central curvature singularity for $0<p<3$. However, for $p=3$, the metric solution
(\ref{solreflat}) is regular with a de Sitter core, while for
$p>3$, the family of solutions are again regular black holes
with an increasingly regular core \cite{Simpson:2019mud}. The region internal to the inner horizon is spacelike and
completely regular at the origin. Setting $p=3$ for definiteness and $2\sigma^2=\pi$ we find that for $\mu_{ext}\sim 1.13$ we have an extremal black hole. For $\mu_{ext}\leq \mu$ we have a sequence of regular black holes whereas for smaller masses than $\mu_{ext}$ we have a regular solution without horizon; spacetime is curved but not sufficiently in
order to create an event horizon. These solutions are gravitational particle-like solutions akin to dark matter, provided they are stable.

We now proceed to scan, starting from weak up to strong gravity, the possible notable differences of our regular solution, as compared to standard GR.
We do not aim to be extensive here, we rather give a first approach that is useful for future studies.
Let us first seek the leading PPN parameters of this solution in order to effectively see how it compares with GR. In order to do this we effectively find a Cartesian distance coordinate $\rho=\sqrt{x^2+y^2+z^2}$ where $(x,y,z)$ are harmonic coordinates suited for a Newtonian gauge. As an example take $p=3$ whereupon we get,
\begin{equation}
r=\rho+M-\frac{4\mu \sigma^2}{\rho^3}+O(1/\rho^4).
\end{equation}
This coordinate system is harmonic for large distances compared to the size of the outer event horizon. Furthermore, to leading order, 
it agrees with the harmonic radial coordinate of Schwarzschild (see \cite{Tucker:2018rgy} for clarification on coordinate issues 
in higher PN calculations). Such distances of the order of some $1400$ Schwarzschild radii correspond to the orbits of stars like S2 orbiting Sgr*A.
Using these coordinates we can quite easily obtain the leading (see for example \cite{Will:2014kxa}) PN parameters, $\beta=\gamma=1$, which end up identical to GR for $p\geq 3$.

We can try to go a step further and evaluate directly the precession of a star like S2 orbiting the massive compact object identified with Sgr A* (see \cite{Abuter:2020dou} and references within). Star S2 orbits the central, regular for our purposes, black hole, following timelike geodesics at the equator $\theta=\pi/2$. Using the Killing
symmetries for rest energy per unit rest mass $E$ and angular momentum per unit rest mass $L$ we have the standard relations,
\begin{equation}
E=h(r)\frac{dt}{d\tau},\qquad L=r^2 \frac{d \phi}{d\tau},
\end{equation}
where $\tau$ is the geodesic parameter.
Transforming to $u=1/r$ coordinates and using the above, it is straightforward to obtain the Binet's modified equation governing the trajectory of $S2$,
\begin{equation}
\frac{d^2 u}{d \phi^2}+u=\frac{\mu }{L^2}(u X_u+ X)+ 3 \mu u^2 X+ \mu u^3 X_u,
\end{equation}
where now $u$ is a function the angular coordinate $\phi$.
The above equation gives us precisely the GR case of Schwarzschild for $X=1$. Binet's original equation, valid for the Newtonian limit, is obtained if we take $X=1$ and we additionally neglect the higher order $3 \mu u^2$ term. This orbital equation is valid for any regular black hole we choose in the face of
$X$ and for classical precession tests of solar system planets. As an example, we can set $p=3$ for our regular solution and Taylor expand for small $u$ (or large $r$),
\begin{equation}
X=1-\frac{4 \sigma^2}{\pi^2} u^3+ O(u^9).
\end{equation}
We get the approximate equation,
\begin{equation}
\frac{d^2 u}{d \phi^2}+u=\frac{\mu }{L^2}+\frac{\epsilon L^2}{ \mu} u^2-\frac{16 \sigma^2 \epsilon}{3\mu \pi^2}u^3+O(u^5).
\end{equation}
Here we have introduced  $\epsilon=\frac{3 \mu^2}{L^2}$ as our small{\footnote{In our geometrized units we have $G=c^2=1$ and therefore $\mu(cm)=0.742 \times10^{-28} \frac{cm}{g} \mu(g)$.}} dimensionless parameter \cite{dinverno:1992}.
We are using the same expansion parameter as for the case of Schwarzschild as we want to point out the difference with the
case of GR. Now expanding $u=u_0+\epsilon u_1$, we obtain to zeroth order the elliptic Kepler trajectory $u_0=\frac{ \mu}{L^2}(1+e \cos \phi)$, where $e$ is the eccentricity. To linear order in $\epsilon$, keeping only the term with growing contribution we find at the end,
\begin{equation}
u \sim \frac{\mu }{L^2} \Big[1+ e \cos[\phi (1-\epsilon f_{SP})]\Big],
\end{equation}
where $f_{SP}=1-8 \frac{\mu \sigma^2 }{L^4 \pi^2}(1+ \frac{e^2}{4})$ denotes our correction
beyond the GR $f_{SP}=1$ value. Constraints from GRAVITY place $f_{SP}\sim 1.1 \pm 0.2$ which
in turn constrains our action parameter $\sigma$. Note however, that given our expansion in
$\epsilon$ we are assuming that our parameter $\sigma^2$ is big enough so as to be of the same order as the Schwarzschild correction. If we adapt our calculation to the orbit characteristics of the S2 star orbit there will be fine-tuning involved.
Generically $f_{SP}=1$ since $\beta=\gamma=1$ for our background.
A similar calculation can be undertaken using null geodesics for time
delay effects akin to pulsars for example (see the review by Johannsen \cite{Johannsen:2015mdd}).

\begin{figure}[h]
\centering
\includegraphics[scale=0.4]{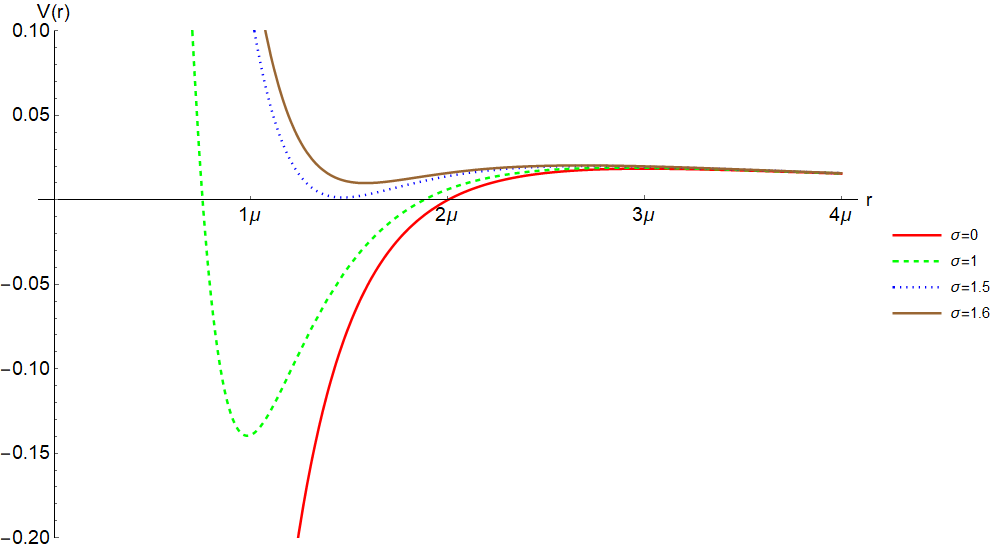}
\caption{Effective potential (\ref{potential}) for different values of $\sigma$ our theory parameter. In particular, $\sigma=0$ corresponds to the effective potential of the Schwarzschild solution for which $X=1$. Varying $\sigma>0$ changes the root of the potential and a non-zero value actually changes the singularity to a minimum. Increasing the value of $\sigma$ further can even remove the root corresponding to the absence of an event horizon altogether. The height of the potential maximum marks $1/b^2_{crit}$ for each curve of the potential.}
\label{effective_potential}
\end{figure}

\begin{figure}[h]
\centering
\includegraphics[scale=0.4]{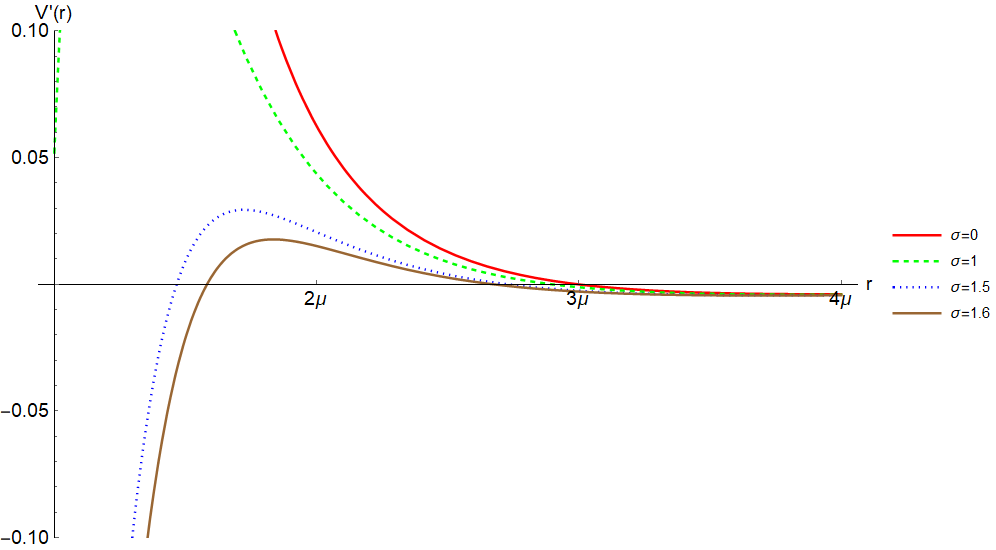}
\caption{Derivative of the effective potential. One can see a small but finite shift of its root, $r_R$, for different values of $\sigma$ as a decreasing function of $\sigma$.}
\label{effective_potential_derivative}
\end{figure}

A last interesting point is to consider our solution in the strong field regime. For our generic purposes
we will pursue here  the light trajectories of photons or massless particles
such as neutrinos in presence of our regular black hole. Again we follow the standard text book
procedure for equatorial geodesics but now we focus on light rays, defining $b=L/E$, the apparent impact parameter, for an observer in the asymptotically flat region. The
parameter $b$ can vary up to  the closest distance photons get to the black hole without being necessarily eaten up by the gravitational well of the black hole. The geodesic equation takes a familiar (particle in a potential) form,
\begin{equation}
\label{ring}
\frac{1}{2}\left( \frac{dr}{d\tilde \tau}\right)^2+\frac{h(r)}{2 r^2}=\frac{1}{2 b^2},
\end{equation}
where we have rescaled $\tilde \tau = L \tau$.
Therefore the effective potential takes the form,
\begin{equation}
\label{potential}
V_{eff}=\frac{1}{2 r^2}\left( 1- \frac{2 \mu}{r} X(r) \right),
\end{equation}
and critical light rings occur at the zeroes of $V'_{eff}=0$ which are the zeroes of the equation,
\begin{equation}
\label{ring1}
r+\mu X' -3 \mu X=0.
\end{equation}
The effective potential and its derivative are depicted in figures \ref{effective_potential} and \ref{effective_potential_derivative} respectively.
Note the familiar light ring solution at $r_R=3\mu$ for Schwarzschild when we set $X=1$. Once we have a zero of (\ref{ring1}), $r=r_R$ we get the maximal impact parameter using (\ref{ring}),
\begin{equation}
b_{crit}=\frac{r_R}{\sqrt{h(r_R)}}.
\end{equation}
The critical impact factor can be as well formulated as
\begin{eqnarray}
b_{crit}=b_{\tiny{\mbox{Schwar.}}} \frac{\left(X(r_R)-\frac{1}{3}X'(r_R)\right)^{\frac{3}{2}}}{\sqrt{X(r_R)-X'(r_R)}}
=b_{\tiny{\mbox{Schwar.}}} \left(\frac{r_R}{3\mu}\right)\sqrt{\frac{4\sigma^4+\pi^2 r_R^6}{\pi^2r_R^6-24\mu r_R\sigma^2+4\sigma^4}},
\end{eqnarray}
where the impact factor for the Schwarzschild solution is given by $b_{\tiny{\mbox{Schwar.}}}=3^{3/2}\mu$. It is easy to see that
$$
\left(\frac{r_R}{3\mu}\right) b_{\tiny{\mbox{Schwar.}}}\leq  b_{crit}\leq \left(\frac{r_R^2}{3\mu}\right)\sqrt{\frac{\pi}{\pi r_R^2-6\mu}}
$$
and the lower bound is achieved for $\sigma=0$ (the Schwarzschild limit) and at the limit $\sigma\to\infty$, corresponding to the flat limit.


The determination of the light ring sets the size of the black hole shadow. The Event Horizon Telescope (EHT) has obtained
the first image of the supermassive M87 black hole. For M87 the size of the shadow was used as a test for GR,
estimating the black hole mass \cite{Akiyama:2019eap}, \cite{Psaltis:2020lvx} and comparing to the independent
calculation for M87's mass given by stellar dynamics \cite{Gebhardt:2011yw}. There are a number of caveats with
this calculation as a test of GR that have primarily to do with the little knowledge of the illuminating accretion
flow for M87 or the sheer mass of the object (see in particular the critical analysis presented in \cite{Glampedakis:2021oie}).
Rather than putting in the numbers we will choose here to sketch the different cases for our regular solution as opposed to Schwarzschild.
For definiteness let us fix the mass of the black hole to $\mu=1$ and vary
the theory parameter $\sigma$ instead, in order to see how the characteristics of the effective potential change as we
sweep through our theory. Indeed we find that for $0<\sigma<\sigma_{ext}$ our effective potential always has a photon ring (outside of the event horizon) and as $\sigma$ is increased we have $r_{R}^\sigma<3$, the GR photon ring case. At the same time, increasing $\sigma$, the height of the potential maximum increases and therefore the critical impact parameter $b_{crit}^\sigma<b_{crit}^0$ is always below the Schwarzschild one
(again see \cite{Glampedakis:2021oie}). Note also that once $\sigma>0$ we always have a minimum of the potential. This scheme continues until we arrive at $\sigma_{ext}$, the case where (for unit mass) we have an extremal black hole. Beyond this point there is no event horizon anymore, for $\mu=1$, and our theories present now two visible critical points, one stable and one unstable.
For a region of impact parameters in between the critical values of the potential, we have bound light orbits for local light sources at $r<3$ or so. This is a distinctive feature of the particle-like solutions and is something that differentiates
them from the regular black hole case. Furthermore, note that photons starting out from infinity can probe into the gravitational solution to all distances. Therefore, for $\sigma>\sigma_{ext}$ there is no longer a central shadow, but rather enhanced light rings very close to the $r=0$ center. In summary, for each given theory (where $p$ and $\sigma$ are fixed) we will have particle-like solutions for $\mu<\mu_{ext}$ and regular black holes for $\mu>\mu_{ext}$.

\section{Thermodynamics of asymptotically flat regular black holes with a scalar field source}

We now turn to the study of the thermodynamic properties of the regular class of black hole solutions (\ref{solreflat}). The thermodynamics of regular solutions is one of the aspects that is widely studied in the literature, see e.g. \cite{Myung:2007xd}. We start by pointing out a difference of our DHOST solution in comparison to regular black holes with non-linear electrodynamics. In the latter case the regularization parameter is actually part of the theory, and is usually associated with a magnetic charge.
This means that the latter solution exists for a fixed value of the magnetic charge, and that to change this value corresponds to changing the theory. A direct consequence of this is that the regularization parameter cannot be considered as a variable parameter, and hence must not appear in the equation of the first law of thermodynamics.
This aspect obscures the thermodynamic interpretation of regular solutions. On the contrary in our case, the regularity of the solution (\ref{solreflat}) is not inherent to the presence of our action bookkeeping parameter $\sigma$, but rather in the presence of the regularizing arctangent function rendering the metric function smooth at the origin. In addition, as it can be seen in Eq. (\ref{solreflat}), the regularizing function comes with a constant $\mu$ which is an integration constant, and hence its interpretation as a thermodynamical variable is not ambiguous.

The thermodynamic analysis of the regular solution (\ref{solreflat}) will be carried out with the Euclidean approach in which the partition function is identified with the Euclidean path integral in the saddle point around the classical solution. In practice, we consider a mini superspace with the following ansatz
\begin{eqnarray}
ds^2=N(r)^2 f(r)d\tau^2+\frac{dr^2}{f(r)}+r^2d\Sigma_{2}^2,\qquad \phi=\phi(r),
\label{EuclAnsatz}
\end{eqnarray}
where $\tau$ (in this section) is the Euclidean (periodic) time with $0<\tau\leq \beta$ and, where
$\beta$ is the inverse of the temperature
\begin{eqnarray}
\beta^{-1}=T=\frac{1}{4\pi}\,N(r) f^{\prime}(r)|_{r_h},
\end{eqnarray}
with $r_h$ being the radius of the horizon. In the mini superspace
defined by the ansatz (\ref{EuclAnsatz}), the Euclidean action $I_E$
(using the proper normalization factor) reads
\begin{eqnarray}
I_E=-\frac{1}{4}\beta\int N\left[\left({\cal P}-2{\cal
Q}^{\prime}\right)f-{\cal Q}\,f^{\prime}+2 G+r^2 K
\right]+B_E, \label{EuclAction}
\end{eqnarray}
where  ${\cal H}, {\cal B}$ and ${\cal
Z}$ are given in (\ref{defHBZ}), and where for simplicity we have defined,
\begin{eqnarray}
{\cal Q}=\frac{\cal B}{4}r^2 X^{\prime} - 2r{\cal H},\qquad {\cal
P}=r X^{\prime}{\cal B}+\frac{r^2}{4}(X^{\prime})^2{\cal Z}-2{\cal
H}. \label{defZa}
\end{eqnarray}
In the Euclidean action (\ref{EuclAction}),  the term
$B_E$ is an appropriate boundary term ensuring that the solution
corresponds to an extremum of the action, and at the same time it codifies  all the
thermodynamic properties. After some algebraic manipulations we get,
\begin{eqnarray}
B_E=\frac{\beta}{4}\lim_{r\to\infty} \left\{\frac{N(r){\cal
Q}(r)X(r)}{r}\right\}\,\mu-\pi\int {\cal Q}(r_h) dr_h.
\end{eqnarray}
On the other hand, since the Euclidean action is related to the Gibbs
free energy ${\cal G}$ through
$$
I_E=\beta\, {\cal G}=\beta\,{\cal M}-{\cal S},
$$
one can easily read off the expressions of the mass ${\cal M}$ and of the entropy ${\cal S}$ from the boundary term,
\begin{eqnarray}
{\cal M}=\frac{1}{4}\lim_{r\to\infty} \left\{\frac{N(r){\cal
Q}(r)X(r)}{r}\right\}\,\mu,\qquad {\cal S}=\pi\int {\cal
Q}(r_h) dr_h. \label{massentropy}
\end{eqnarray}
For the specific regular black hole solution (\ref{solreflat}), these expressions reduce to
\begin{eqnarray}
{\cal M}=\frac{1}{6}\frac{r_h}{\arctan\left(\frac{1}{2}\pi r_h^p \sigma^{1-p}\right)}, \qquad {\cal S}=
 \frac{2}{3} \int \frac{\pi  r_h}{\arctan\left(\frac{1}{2}\pi r_h^p \sigma^{1-p}\right)} dr_h,
\end{eqnarray}
while the temperature is given by
\begin{eqnarray*}
T=\frac{1}{4 \pi  r_h}\left(1-\frac{2\pi\sigma^{p-1} p r_h^p}{\left(\pi^2 r_h^{2 p}+4\sigma^{2p-2}\right) \arctan\left(\frac{1}{2}\pi r_h^p \sigma^{1-p}\right)}\right).
\end{eqnarray*}
It is clear from these relations that the mass and the entropy of the regular solution are positive, and although we do not have a closed form of the entropy we can nonetheless verify the validity of the first law $d{\cal M}=T\,d{\cal S}$. We also note that the entropy of the regular solution does not satisfy
the area law. In fact, from the generic expression as obtained in (\ref{massentropy}), the only way  for the entropy to
satisfy the area law is that the function ${\cal Q}$, as
defined in (\ref{defZa}), must be proportional to ${\cal Q}(r)\propto r$. However, it is a simple matter to check that the solutions of the
field equations given by (\ref{seed3}), and for an ansatz of the form
(\ref{ansatzz}) will necessarily imply that
$$
{\cal Q}(r)\propto \frac{r}{X(r)},
$$
and, consequently the entropy will be proportional to one-quarter of the area only for a
constant kinetic term. On the other hand, our analysis shows that a constant kinetic term is incompatible with the regularity of the solution. Hence, we deduce that for the DHOST theories considered here the regularity of the solutions fitting our ansatz (\ref{ansatzz}) will not be compatible with the one-quarter area law for the entropy. This is not uncommon for modified gravity theories and is understood geometrically in certain cases such as Einstein-Gauss-Bonnet theory (see for example \cite{Charmousis:2008kc}).

\begin{figure}[h]
\centering
\includegraphics[scale=0.4]{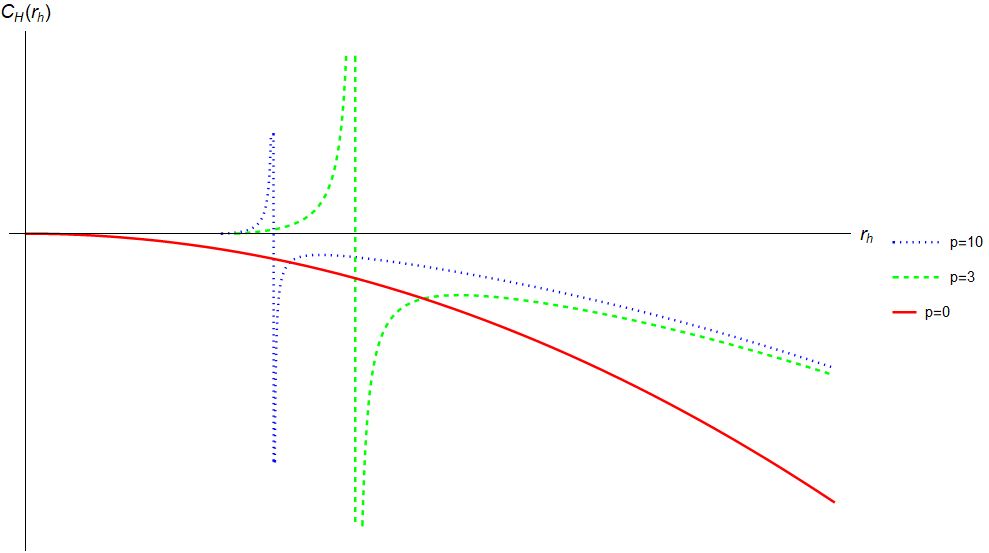}
\caption{Heat capacity of the (\ref{flatdhost}) black hole for different values of $p$ and $\sigma$ such that $2\sigma^{p-1}=\pi$ starting at $r_{\mathrm{Extremal}}$ respectively. Note that these correspond to different theories. There is a second order phase transition at $r_{\mathrm{PT}}$. The asymptotic behavior is like $\propto -r^2$ at infinity. Setting $p=0$ corresponds to the Schwarzschild solution, which has no phase transition.}
\label{heat_capacity_flatdhost}
\end{figure}

Thermodynamic stability of the regular solution is addressed  by computing the heat capacity $C_H=T\frac{\partial{\cal S}}{\partial T}$. From this definition it becomes clear that the heat capacity will provide information about the thermal stability with respect to
the temperature fluctuations, and that a positive heat capacity is a necessary condition to ensure the local stability of the system.
Also, the critical hypersurfaces, that is those where $C_H$ vanishes or diverges, will correspond to the extrema of the temperature with respect to the entropy. For technical reasons, it is more convenient to express the heat capacity as
$$
C_H=T\frac{\partial{\cal S}}{\partial T}=T\left(\frac{\partial{\cal S} }{\partial r_h}\right)\left(\frac{\partial T}{\partial r_h}\right)^{-1},
$$
and, for the regular black hole solution (\ref{solreflat}) we get
\begin{eqnarray}
C_{\mathrm H} = \frac{2 \pi  C r_h^2 \left(r_h^{2 p}+\frac{4}{\pi^2}\sigma^{2p-2}\right) \left[\left(r_h^{2 p}+\frac{4}{\pi^2}\sigma^{2p-2}\right) \arctan\left(\frac{1}{2}\pi r_h^p \sigma^{1-p}\right)- \frac{2}{\pi}\sigma^{p-1} p r_h^p\right]}{\mathcal{C}}, \nonumber \qquad
\end{eqnarray}
with
{\small
\begin{eqnarray*}
\mathcal{C}&=&3 \arctan\left(\frac{1}{2}\pi r_h^p \sigma^{1-p}\right) \left[ \frac{2}{\pi}\sigma^{p-1} p \left( \frac{4}{\pi^2}\sigma^{2p-2} (p-1) -(p+1) r_h^{2 p} \right) r_h^p+\left(r_h^{2 p}+\frac{4}{\pi^2}\sigma^{2p-2}\right)^2 \arctan\left(\frac{1}{2}\pi r_h^p \sigma^{1-p}\right)\right] \\
&&- \frac{12}{\pi^2}\sigma^{2p-2} p^2 r_h^{2 p}.
\end{eqnarray*}}
Due to its lengthy form it is insightful to plot the heat capacities. The heat capacities are shown in figure \ref{heat_capacity_flatdhost}, where
we have excluded the part that corresponds to negative temperatures (akin to the presence of an internal horizon). From this picture, one can see that only small black holes are locally stable  and a critical hypersurface will emerge at some positive radius revealing the existence of a second order
phase transition, as it is the case for the non-linear electrodynamical regular black holes, see e. g. \cite{Myung:2007xd}.

\begin{figure}[h]
\centering
\includegraphics[scale=0.4]{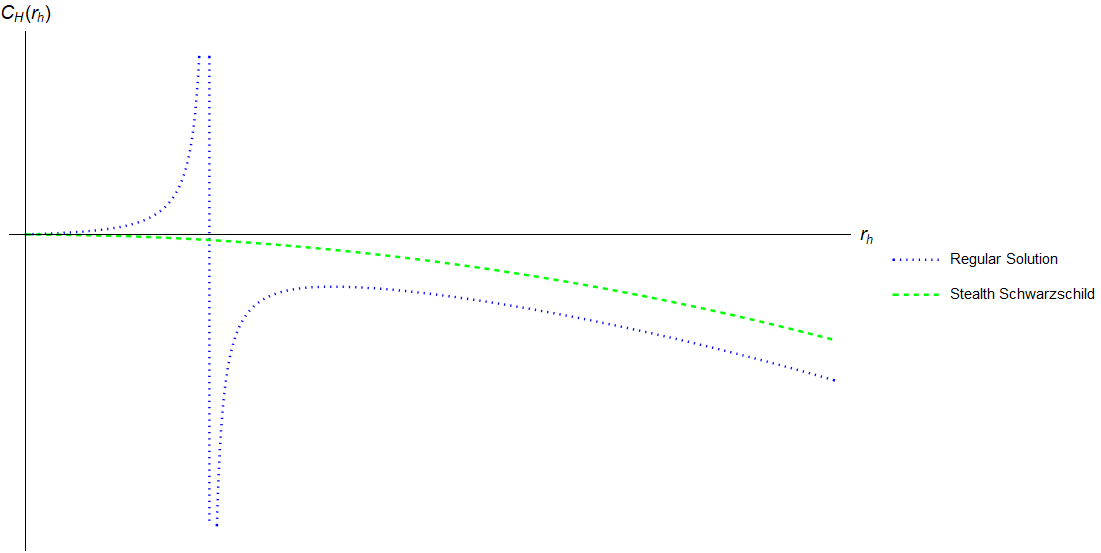}
\caption{Heat capacity of the (\ref{flatdhost}) black hole for $p=1$ and the stealth Schwarzschild solution. This time they correspond to the same theories, even though their behaviour looks identical to before. Further the temperature is positive everywhere, so there is no extremal value of $r$ and the heat capacities can be plotted from $r=0$.}
\label{heat_capacity_stealth}
\end{figure}

Before closing this section, we would like to address the following question: for the DHOST theory as defined in Appendix,
does there exist another solution, and if so, would this allow for a thermodynamic stability comparison of the two solutions? In order to answer this question, we notice that the first equation (\ref{seed1}) gives,
\begin{eqnarray}
0 &=& \frac{16 \left[ \frac{2}{\pi}\sigma^{p-1} \sin\left( \frac{\pi}{2}X \right) \right]^{-\frac{2}{p}}}{3\pi^2\left[ \frac{2}{\pi} p \sin\left( \frac{\pi}{2}X \right) - 6 X \right]^4 X} \left[ - r^2 \cos\left( \frac{\pi}{2}X \right)^{\frac{2}{p}} + \left( \frac{2}{\pi}\sigma^{p-1} \sin\left( \frac{\pi}{2}X \right) \right)^{\frac{2}{p}} \right] F[X],
\end{eqnarray}
with $F[X]$ being an algebraic equation in $X$ given by
\begin{eqnarray*}
F[X] &=& 72 X^2 \left[p^2 \cos \left(2\pi X\right)-p \cos \left(\pi X\right)-2\right]- \frac{32}{\pi^2} p^2 \sin ^2\left(\pi X\right) \left[p \cos \left(\pi X\right)-4\right] \\
&+& \frac{12}{\pi} p X \sin \left(\pi X\right) \left[p^2 \cos \left(2\pi X\right)+3 p^2-26 p \cos \left(\pi X\right)+26\right].
\end{eqnarray*}
From this it is easy to see that there are only two possibilities: either $X$ is given by the previous form (\ref{solreflat}), or $X$ is a constant solving
the constraint $F[X]=0$. On the other hand, taking the difference between (\ref{seed2}-\ref{seed4}) yields $f(r) = h(r)$,
so in the first case we end up with the regular black hole. After some straightforward computations, we can establish
that only the DHOST theory defined in the Appendix with $p=1$ will admit two different solutions, and one of these is a
stealth Schwarzschild black hole configuration given by
\begin{eqnarray}
\label{stealthsol}
h(r)=f(r) = 1 - \frac{\mu}{r}, \qquad X= 1 + 2 n ,
\end{eqnarray}
where $n$ is an integer number. The thermodynamic quantities of this stealth solution are given by
\begin{eqnarray}
{\cal M}= \frac{r_h}{3\pi},\quad {\cal S}=\frac{2}{3} r_h^2,\quad
T=\frac{1}{4\pi r_h}, \quad C_{\mathrm H} = -\frac{4}{3}r_h^2,
\end{eqnarray}
and as stressed before the entropy satisfies the area law
because of the constant value of the kinetic term (\ref{stealthsol}). The comparison of the respective heat capacities can be seen in Figure \ref{heat_capacity_stealth}.
We can now compare the $\arctan-$solution
(\ref{solreflat}) for $p=1$ with the stealth solution (\ref{stealthsol}). Using the free energy, defined as ${\cal F}= {\cal M} - T {\cal S}$, one can calculate the difference of the respective solutions at equal temperatures
\begin{eqnarray*}
\Delta {\cal F} &=& F_{\mathrm{regular}} - F_{\mathrm{stealth}} = T \int {\cal F}(r_h) dr_h, \\
{\cal F}(r) &=& \frac{r \left[-4 \left(r^2+1\right) \arctan(r)^2+\pi  \left(r^2+1\right) \arctan(r)-\pi  r\right] \left[-2 r^3 \arctan(r)-r^2+\left(r^2+1\right)^2 \arctan(r)^2\right]}{\arctan(r) \left[\left(r^2+1\right) \arctan(r)-r\right]^3}
\end{eqnarray*}
It is easy to notice that the integrand ${\cal F}(r)$, goes to $+\infty$ for $r \rightarrow 0$ and to $-\infty$ for $r \rightarrow \infty$. Hence,
one would expect the stealth solution to be thermodynamically favored for small $r_h$, and there is the possibility
that this changes for sufficiently large $r_h$. However, because of its lengthy integral form it is not
possible to make any exact statements about this.

\section{Conclusions}
Making use of a generalized Kerr-Schild solution generating method, as described in \cite{Babichev:2020qpr}, we have constructed a family of regular black holes, namely solutions without curvature singularities. They are characterized by the presence of an arctangent regularizing function, and are regular solutions of specific higher-order
scalar tensor theories known as DHOST theories. The solutions are asymptotically flat and are accompanied by a regular scalar field. They are characterized by a de Sitter or,  increasingly regular core, inner and outer event horizons and particle-like regular solutions. The latter appear depending on a certain theory strength parameter $\sigma$ (related to the mass) and could have a distinct phenomenology as compared to black holes due to the absence of the horizon. Indeed we examined a number of observable consequences of our solutions ranging from weaker to stronger gravity: from the leading post-Newtonnian Eddington parameters to leading precession effects up to enhanced geodesic light rings. It would be interesting to go beyond our initial calculations and check for example echoes of our particle-like solutions as predicted in \cite{Cardoso:2016rao}. Very recent similar studies have shown such effects in the case of Einstein-Gauss-Bonnet theories \cite{Kleihaus:2020qwo} and it would be interesting to apply known methods for our analytic explicit solutions.

Our regular black hole solutions differ from existing models
of regular solutions in several ways. First of all, it is important to stress that the
DHOST models for which regular black holes exist are not finetuned by some regularizing parameter, which is usually the case for regular black holes. Regularity of the solution is achieved directly by the form of the kinetic $X(r)$ function. As a direct
consequence the regular solutions (once regularity of the core is fixed) only depend on a unique integration constant, mass and a bookkeeping parameter $\sigma$ which measures the magnitude of the higher order effects (the limiting case $\sigma\to 0$ gives GR).
This is a major difference with respect to the regular black holes
of non-linear electrodynamic models, since in those
cases the mass, as well as the regularizing parameter (usually
associated to a magnetic charge), are part of the non-linear
electrodynamic Lagrangian. In the
present case, the regular solutions only depend on a unique
integration constant, which is shown to be proportional to the mass. We also note that the "usual" area law for the entropy is not compatible with the regularity of
our solution (\ref{ansatzz}-\ref{massfunction}-\ref{hyp1}) and this is due to the theory's modified nature of gravity. This is quite common and understood in certain cases due to the higher order nature of the theory (see for example \cite{Charmousis:2008kc}).
In spite of the violation of the  area law, we have shown
that the first law of thermodynamics is always satisfied. The
regular black hole solutions have a mass fall-off of the form $\frac{\arctan(r^p)}{r}$, where $p>0$ is a parameter of the theory. Note that examples of black hole solutions with such regular terms at the origin have been encountered \cite{Babichev:2017rti} as AdS solitons. We have seen that the small regular black holes are thermodynamically stable since their heat capacity turns out to be positive and for the range of values of the
parameter ensuring the regularity solution, we have observed the existence of second order phase transitions for all our regular black holes.

It would be interesting to question if regularity of such solutions in DHOST theories persists once these are rotating. Given the recent progress in this direction \cite{Anson:2020trg} there may be hope in such a direction, even analytically. Furthermore, it would be an interesting first step to extend regular solutions to the presence of a time dependent scalar field in order to understand how the picture of geodesics is altered with regularity. These are some of the possible directions in this exciting field that we hope to pursue in the near future.

\section*{Acknowledgements}
We would like to thank Tim Anson, Eloy Ay\'on-Beato, Eugeny Babichev, Thanasis Bakopoulos, Alessandro Fabbri, Panagiota Kanti, Antoine Leh\'ebel  and Georgios Pappas for many enlightening
discussions. The authors also gratefully acknowledge the kind
support of the PROGRAMA DE COOPERACI\'ON CIENT\'IFICA ECOSud-CONICYT
180011/C18U04. OB is funded by the PhD scholarship of the University of Talca. The work of MH has been partially supported by FONDECYT grant 1210889. The work of MSJ is funded by the National Agency for Research and Development (ANID) / Scholarship Program/ DOCTORADO BECA NACIONAL/ 2019 - 21192009

\section*{Appendix: DHOST models for the regular solution (\ref{solreflat})}
Along the lines of \cite{Babichev:2020qpr}, one can show that the DHOST action defined by
{\small
\begin{eqnarray*}
&& {\cal H}(X)=-\frac{2}{3\pi X - p\sin(\pi X)},\\
&&G(X)=\frac{p^2\sin(2\pi X)-8 p\sin(\pi X)+6\pi X}{\left(p\sin(\pi X)-3\pi X\right)^2},\\
&& A_1(X)=\frac{2 p\sin(\pi X)(p\cos(\pi X) - 3)}{X(p\sin(\pi X) - 3\pi X)^2}, \\
&&K(X)=\frac{p\sin(\frac{\pi}{2}X)^{\frac{p - 2}{p}}\cos(\frac{\pi}{2}X)^{\frac{p + 2}{p}}\left (B^2 p^2\cos(2\pi X) - B^2p^2 - 24pX^2\cos(\pi X) + 28BpX\sin(\pi X) - 24X^2\right)}{3X^2A^\frac{2}{p}(p\sin(\pi X) - 3\pi X)^2},
\end{eqnarray*}}
and
{\tiny
$$
A_3(X)=\frac{B\left(2p^2\left(5B^2 + 144X^2\right)\cos(2\pi X) + 3p\left(B^2p^2 - 192X^2\right)\cos(\pi X) - 3B^2p^3\cos(3\pi X) - 10B^2p^2 + 24BpX\sin(\pi X)\left(-23p\cos(\pi X) + 2p^2 + 43\right) - 288X^2\right)}{3X^2\left(Bp\sin(\pi X) - 6X\right)^3} , \\
$$}
where $A=\frac{2\sigma^{p-1}}{\pi}$ and $B=\frac{2}{\pi}$ and $\sigma$ an unspecified constant, admits the following regular black hole solution
\begin{eqnarray}
&&ds^2=-\left(1-\frac{2\mu \arctan\left(\frac{\pi r^p}{2 \sigma^{p-1}}\right)}{\pi r}\right)dt^2+\frac{dr^2}{\left(1-\frac{2\mu \arctan\left(\frac{\pi r^p}{2 \sigma^{p-1}}\right)}{\pi r}\right)}+r^2(d\theta^2+\sin^2\theta
d\varphi^2), \nonumber \\
&&X(r)= \frac{2}{\pi}\arctan\left(\frac{\pi r^p}{2 \sigma^{p-1}}\right).
\label{flatdhost}
\end{eqnarray}

\end{document}